\documentclass[aps, prl, reprint, twocolumn,superscriptaddress]{revtex4-1}

\usepackage{graphicx}
\usepackage{bm}
\usepackage{amsmath}
\usepackage{amssymb}
\newcommand{\um}[0]{\mu\mathrm{m}}

\newcommand{\rmi}{{\rm i}}
\newcommand {\e}{{\rm e}}

\newcommand{\fSAW}[0]{f_\mathrm{SAW}}
\newcommand{\lSAW}[0]{\lambda_\mathrm{SAW}}

\newcommand{\sleft}[0]{S_\mathrm{11}}
\newcommand{\sright}[0]{S_\mathrm{22}}
\newcommand{\st}[0]{S_\mathrm{21}}
\newcommand{\epsxx}[0]{u_{xx}}
\newcommand{\epsxz}[0]{u_{xz}}
\newcommand{\epszz}[0]{u_{zz}}

\newcommand{\vBext}[0]{\mathbf{B}}
\newcommand{\Bext}[0]{B}
\newcommand{\Bangle}[0]{\phi_\mathrm{B}}
\newcommand{\Dm}[0]{\Delta m_S}
\newcommand{\mS}[0]{m_S}

\newcommand{\VSi}[0]{\mathrm{V_{Si}}}

\begin{document}

\title{Anisotropic Spin-Acoustic Resonance in Silicon Carbide at Room Temperature}
\author{A.~Hern\'{a}ndez-M\'{i}nguez}
\email{alberto.h.minguez@pdi-berlin.de}
\affiliation{Paul-Drude-Institut f\"{u}r Festk\"{o}rperelektronik, Leibniz-Institut im Forschungsverbund Berlin e.V., Hausvogteiplatz 5-7, 10117 Berlin, Germany}
\author{A.~V.~Poshakinskiy}
\affiliation{Ioffe Physical-Technical Institute, Russian Academy of Sciences, 194021 St.~Petersburg, Russia}
\author{M.~Hollenbach}
\affiliation{Helmholtz-Zentrum Dresden-Rossendorf, Institute of Ion Beam Physics and Materials Research, Bautzner Landstrasse 400, 01328 Dresden, Germany}
\affiliation{Technische Universit\"{a}t Dresden, 01062 Dresden, Germany}
\author{P.~V.~Santos}
\affiliation{Paul-Drude-Institut f\"{u}r Festk\"{o}rperelektronik, Leibniz-Institut im Forschungsverbund Berlin e.V., Hausvogteiplatz 5-7, 10117 Berlin, Germany}
\author{G.~V.~Astakhov}
\affiliation{Helmholtz-Zentrum Dresden-Rossendorf, Institute of Ion Beam Physics and Materials Research, Bautzner Landstrasse 400, 01328 Dresden, Germany}

\date{\today}

\begin{abstract}
We report on acoustically driven spin resonances in atomic-scale centers in silicon carbide at room temperature. Specifically, we use a surface acoustic wave cavity to selectively address spin transitions with magnetic quantum number differences of $\pm 1$ and $\pm 2$ in the absence of external microwave electromagnetic fields. These spin-acoustic resonances reveal a non-trivial dependence on the static magnetic field orientation, which is attributed to the intrinsic symmetry of the acoustic fields combined with the peculiar properties of a half-integer spin system. We develop a microscopic model of the spin-acoustic interaction, which describes our experimental data without fitting parameters. Furthermore, we predict that traveling surface waves lead to a chiral spin-acoustic resonance, which changes upon magnetic field inversion. These results establish silicon carbide as a highly-promising hybrid platform for on-chip spin-optomechanical quantum control enabling engineered interactions at room temperature. 
\end{abstract}

\maketitle


Hybrid spin-mechanical systems are considered as a promising platform for the implementation of universal quantum transducers \cite{Schuetz:2015dx} and ultra-sensitive quantum sensors \cite{Poshakinskiy:2019bi}. Spin states can be coupled by the strain fields of phonons and mechanical vibrations. Coherent sensing of mechanical resonators \cite{Kolkowitz:2012iw}, acoustic control of single spins \cite{Maity:2020cn} and electromechanical stabilization of spin color centers \cite{Machielse:2019bt} based on spin-optomechanical coupling have been demonstrated. Similarly to a magnetic field, the application of a static strain field leads to a shift of the spin levels, while a resonantly oscillating strain field induces interlevel spin transitions. Their selection rules are imprinted by the crystal symmetry or device geometry, which provide a high degree of flexibility for on-chip coherent spin manipulation \cite{Lee:2017gs, Satzinger:2018et,Takada:2019cq} and may support chiral spin-phonon coupling \cite{Zhu:2018fs}. 

Most of the systems coupling atomic-scale spins to vibrations studied so far are based on color centers in diamond \cite{MacQuarrie:2013cp, Arcizet:2011cg, Kolkowitz:2012iw, Teissier:2014gt,  Golter:2016cd, Chen:2019ks, Machielse:2019bt, Maity:2020cn}. Two characteristics of silicon carbide (SiC) make it a natural material choice for hybrid spin optomechanics. As diamond, SiC hosts highly-coherent optically-active spin centers, such as negatively charged silicon vacancies ($\VSi$) \cite{Riedel:2012jq} and divacancies (VV) \cite{Falk:2013jq}. In addition, SiC is already used in commercial nano-electro-mechanical systems (NEMS) with robust performance and ultrahigh sensitivity to vibrations~\cite{Li:2007ex}. Recently, the mechanical tuning \cite{Falk:2014fh} and acoustic coherent control \cite{Whiteley:2019eu} of the VV spin $S=1$ centers in SiC have been demonstrated at cryogenic temperatures. However, symmetry-dependent spin-acoustic interactions are still largely unexplored and SiC-based hybrid spin-mechanical systems under ambient conditions remain elusive.

In this letter, we demonstrate room-temperature spin-acoustic resonance (SAR) in 4H-SiC. 
We exploit the intrinsic properties of the half-integer spin $S=3/2$ system \cite{Kraus:2013di}, the so-called $\VSi$ qudit \cite{Soltamov:2019hr}, to realize full control of the spin states using high-frequency vibrations. This is fulfilled by acoustically coupling spin sublevels with magnetic quantum numbers ($\mS$) differing by both  $\Dm =\pm 1$ and $\Dm = \pm 2$. In contrast to previous SAR studies, which were restricted to $S=1$ atomic-scale spins \cite{MacQuarrie:2013cp, Arcizet:2011cg, Kolkowitz:2012iw, Teissier:2014gt,  Golter:2016cd, Chen:2019ks, Machielse:2019bt, Maity:2020cn,Falk:2014fh,Whiteley:2019eu}, the spin $S=3/2$ system enables all-optical readout of the spin state without application of microwave electromagnetic fields. The mechanical vibrations are applied via a surface acoustic wave (SAW) resonator patterned on the 4H-SiC wafer surface perpendicular to the c-axis. The superposition of axial and shear strain components with different amplitudes and phases makes the SARs dependent on the quantization direction of the spin states relative to the SAW propagation direction. Here, this anisotropy manifests itself as a complex angular dependence of the $\Dm = \pm 1$ and $\Dm = \pm 2$ transitions on the orientation of the external magnetic field $\vBext = (B_x,B_y, 0)$ applied in the plane perpendicular to the c-axis of 4H-SiC (i.e., with $B_z = 0$). We develop a model for spin-3/2 SARs using an effective spin-strain coupling Hamiltonian~\cite{Poshakinskiy:2019bi, Udvarhelyi:2018tg}, which can describe our non-trivial observations. The selective excitation of transitions with  $\Dm =\pm 1$ and $\Dm = \pm 2$ as well as the anisotropic behavior with magnetic field orientation demonstrated here provides extra degrees of freedom for the control of interactions in atomic-scale spin systems.

\begin{figure}
\includegraphics[width=\linewidth]{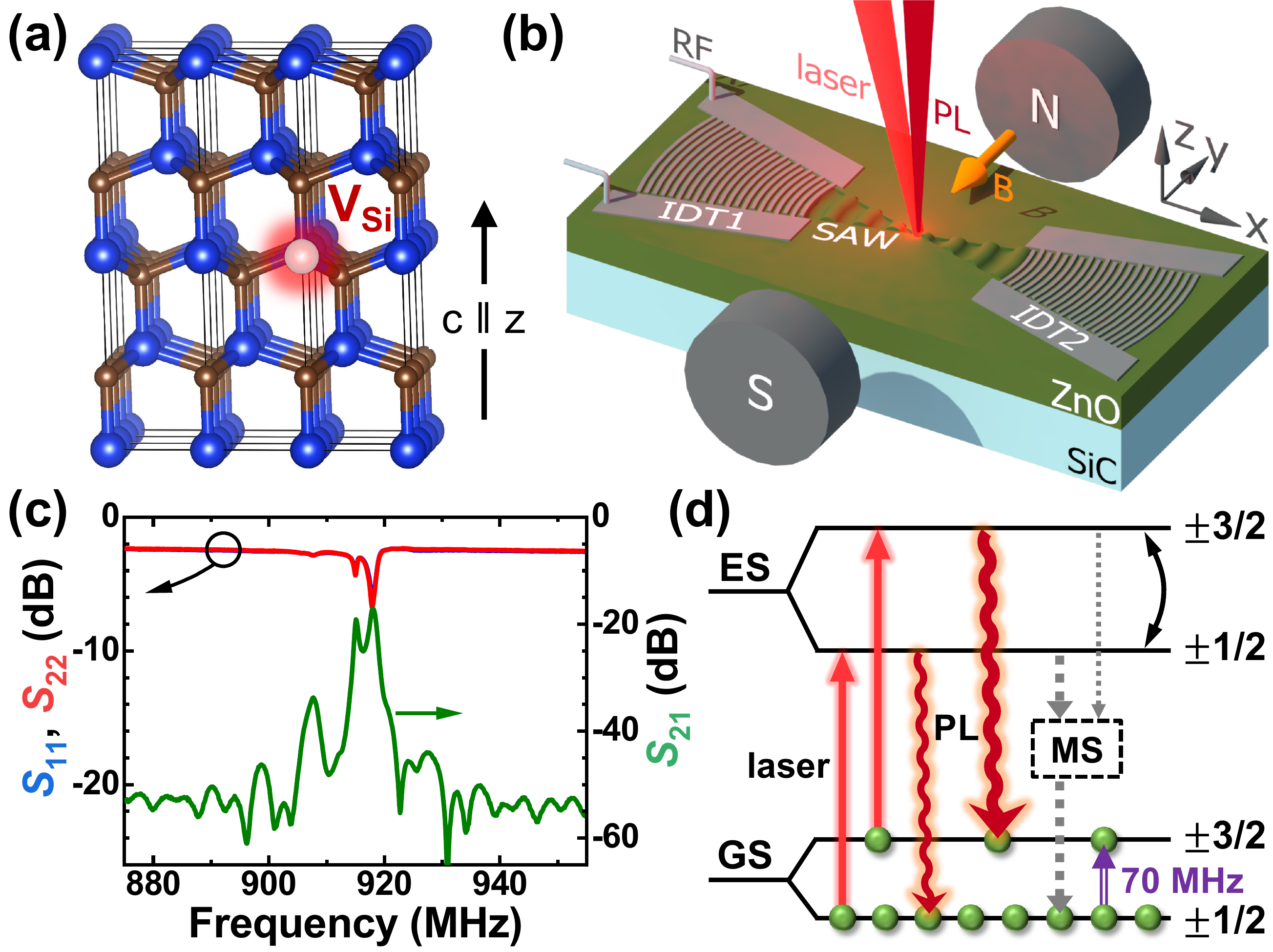}
\caption{Schematic presentation of (a) the 4H-SiC lattice with one $\VSi$ defect and (b) the acoustic device. Focusing IDTs excite and detect SAWs propagating in the ZnO-coated SiC. The sample is placed between the poles of an electromagnet for the application of a magnetic field $\vBext$ in the plane of the sample surface. (c) RF scattering parameters for IDT1 and IDT2. $\sleft$ and $\sright$ correspond to the power reflection coefficient of IDT1 and IDT2, respectively, and  $\st$ to the power transmission coefficient. (d) Optical pumping cycle and readout principle of the $\VSi$ spin. The double vertical arrow indicates the spin resonance transition at 70~MHz in zero magnetic field (see Supplemental Material \cite{Note2}).}
\label{fig1}
\end{figure}

\begin{figure}
\includegraphics[width=\linewidth]{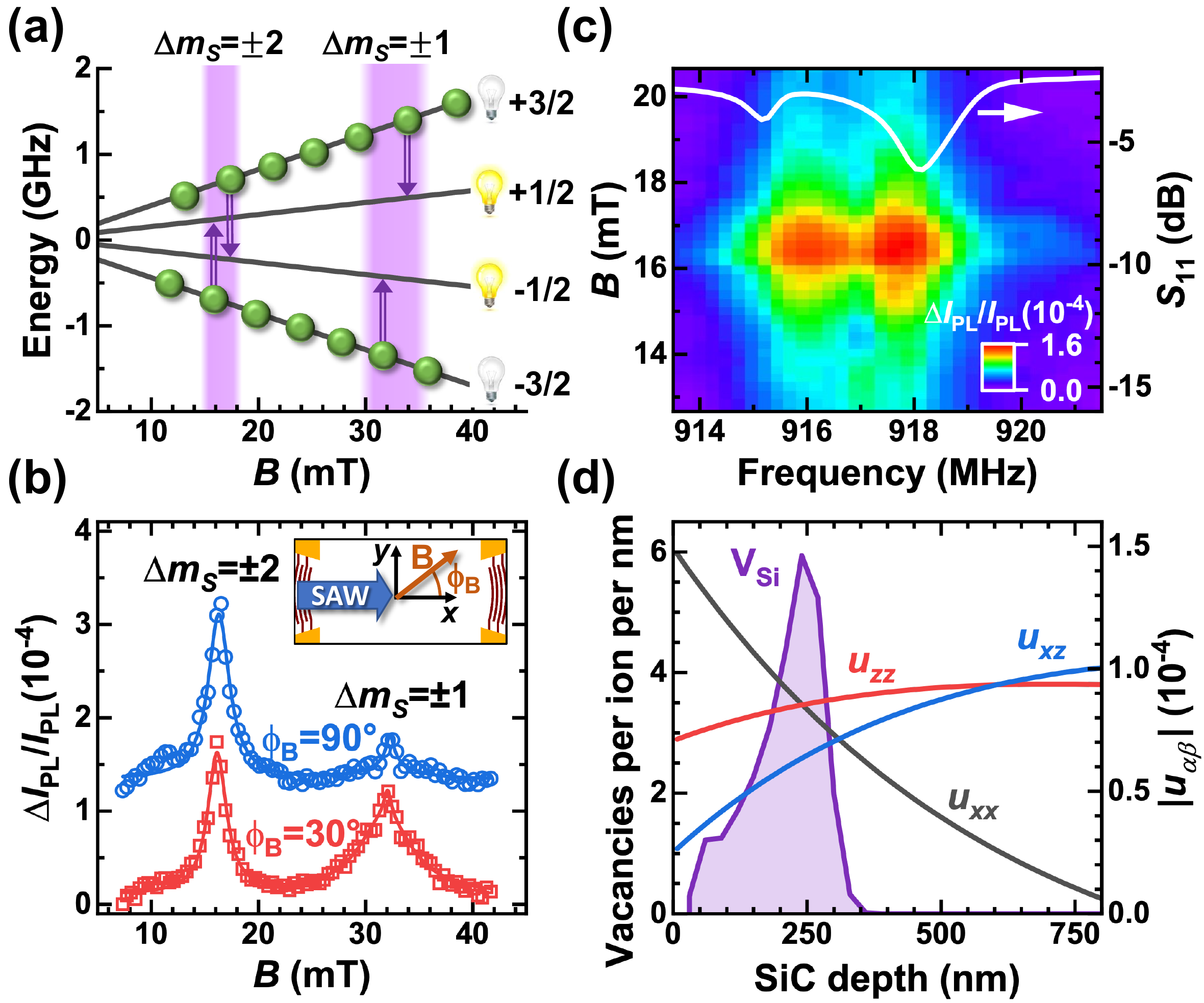}
\caption{(a) Evolution of the spin levels in the $\VSi$ GS under application of the in-plane magnetic field $\vBext\perp c$. The vertical double arrows indicate the resonant spin transitions at $916 \, \mathrm{MHz}$ induced by the SAW. (b)  The SAR signal as a function of the magnetic field strength. The data are measured for two magnetic field angles ($\Bangle$) with respect to the SAW propagation direction (see inset). The solid curves are fits to a multi-peak Lorentzial function. The data are vertically shifted for clarity (c) Two-dimensional color map of the $\Dm=\pm2$ spin transition as a function of the magnetic field strength and the RF frequency applied to the IDT, and power reflection coefficient $\sleft$ of the IDT. (d) Simulated depth profiles for the distribution of vacancies after irradiation (which is proportional to the density of $\VSi$ centers) and of the $\epsxx$, $\epszz$ and $\epsxz$ strain amplitudes.}
\label{fig2}
\end{figure}

We present results for the V2 center, corresponding to one of the two possible $\VSi$ crystallographic configurations  \cite{Ivady:2017bq}. Figure~\ref{fig1}(a) displays the 4H-SiC lattice with a single $\VSi$ center.  The $\VSi$ centers are created in a $10\times10$~mm$^2$ semi-insulating 4H-SiC substrate by the irradiation with protons with an energy of 37.5~keV to a fluence of $10^{15}$~cm$^{-2}$ \cite{Kraus:2017cka}. Figure~\ref{fig2}(d) shows the calculated depth distribution of the $\VSi$ centers, which has a mean depth of 250~nm below the SiC surface \footnote{The depth distribution was determined using SRIM (stopping and range of ions in matter) simulation~\cite{SRIM}.}. As shown in the Supplemental Material (SM) \footnote{See Supplemental Material at URL for photoluminescence characterization, ODMR spectrum at zero magnetic field, spatial dependence of the SAR, dependence of the SAR on SAW power, and determination of the matrix elements for the SAW-induced spin transition.}, these centers reveal the 70~MHz zero-field spin splitting characteristic of the V2 centers~\cite{Soerman2000, Bardeleben2000, Orlinski2003, Tarasenko_pssb255_1700258_2018}. After irradiation, the SiC substrate is coated with a 35-nm-thick SiO$_2$ layer followed by a 700-nm-thick ZnO piezoelectric film using radio-frequency (RF) magnetron sputtering. Finally, acoustic cavities defined by a pair of focusing interdigital transducers (IDTs) are patterned on the surface of the ZnO film by electron beam lithography and metal evaporation. Figure~\ref{fig1}(b) displays a schematic representation of our acoustic device. Each IDT consists of 80 aluminum finger pairs for excitation/detection of SAWs with a wavelength $\lSAW=6~\um$, and an additional Bragg reflector consisting of 40 finger pairs placed on its back side (not shown in Fig.~\ref{fig1}). The finger curvature and separation between the opposite IDTs ($\approx 120~\um$) are designed to focus the SAW beam at the center of the cavity. Figure~\ref{fig1}(c) displays the RF scattering ($S$) parameters of the IDTs measured with a vector network analyzer. They show a series of sharp dips within the resonance band of the IDT at a frequency $\fSAW\approx916$~MHz, which correspond to the excitation of the Rayleigh SAW modes of the resonator. 

Figure~\ref{fig1}(d) displays a simplified energy diagram of the V2 $\VSi$ center  \cite{Soltamov:2015ez, Ivady:2017bq} together with the optical pumping and readout scheme~\cite{Simin:2016cp}. The $\VSi$ center has spin 3/2, which is split in zero magnetic field into two Kramer's doublets due to  a low symmetry of the $\VSi$ center ($C_{3v}$ point group). The zero-field splitting between the $\mS=\pm1/2$ and $\mS=\pm3/2$ doublets is equal to $70$~MHz, with the spin quantized along the $c$-axis. Optical excitation into the excited state (ES), followed by spin-dependent recombination via the metastable state (MS), leads to a preferential population of the $\mS=\pm1/2$ spin sublevels in the ground state (GS), as indicated by the green dots in Fig.~\ref{fig1}(d). As the photoluminescence (PL) intensity is stronger for the $\mS=\pm3/2$ states, the PL is sensitive to the resonant spin transitions between the $\mS=\pm1/2$ and $\mS=\pm3/2$ sublevels (see~SM \cite{Note2})~\cite{Tarasenko_pssb255_1700258_2018}. 

The optically detected SAR experiments are performed in a confocal micro-photoluminescence ($\mu$-PL) setup, as illustrated in Fig~\ref{fig1}(b). The SAWs are generated by applying  to one of the IDTs an amplitude-modulated RF signal of appropriate frequency. The sample is excited by a Ti-Sapphire laser (at a wavelength of 780~nm) focused onto a spot size of $10~\mu$m. The  $\VSi$ PL band centered around 900~nm (see SM \cite{Note2}) is collected by an objective, spectrally filtered and detected by a photodiode detector connected to an amplifier locked-in to the RF modulation frequency. The GS spin transition frequencies are tuned to the SAW resonance frequency by applying the in-plane magnetic field $\vBext$. 

To describe the spin-acoustic interaction of the $\VSi$ center in an external magnetic field,  we consider an effective spin-3/2 Hamiltonian 

\begin{eqnarray}\label{eq_Hamiltonian}
\mathcal{H}&=&\mathcal{H}_B+\mathcal{H}_{\rm def}\text{, with}\\
\mathcal{H}_B&=&g\mu_B\mathbf{S}\cdot\mathbf{B}+D\left(S_z^2-\frac{5}{4}\right).
\end{eqnarray}

\noindent Here, $\mathbf{S}=(S_x, S_y, S_z)$ is the spin-3/2 operator, $\mu_B$ is the Bohr magneton, $g\approx2$  the in-plane $g$-factor, and $D = 35$~MHz  the zero-field splitting constant. $\mathcal{H}_B$ describes the Zeeman splitting in $\vBext=(B_x, B_y, 0)$. For $\vBext=0$, this Hamiltonian yields the GS eigenstates displayed in Fig.~\ref{fig1}(d). As discussed below,  $\mathcal{H}_{\rm def}$ describes the coupling of the $\VSi$ spin and elastic deformations \cite{Poshakinskiy:2019bi, Udvarhelyi:2018tg}.   

Figure~\ref{fig2}(a) shows the Zeeman shift of the GS spin sublevels calculated from $\mathcal{H}_B$. For $B \gtrsim 2.5 $~mT, the Zeeman splitting in the GS is larger than the zero-field splitting. Then, the spin quantization axis is along the $\vBext$ direction and the spin sublevels shift linearly with the magnetic field. We note that for in-plane magnetic fields, the states with spin projection on magnetic field direction $\mS=\pm3/2$ are preferentially populated under optical pumping (as indicated by the green dots). In contrast to the $\vBext=0$ case illustrated in Fig.~\ref{fig1}(c), the PL is now stronger for transitions between the ES and GS involving the $\mS=\pm1/2$ states (see the light bulbs next to each energy level).

The optically detected SAR as a function of  $\Bext$ is presented in Fig.~\ref{fig2}(b). We observe two resonances at $\Bext =16.7$~mT and  $\Bext = 33.3$~mT, which are ascribed to the acoustically driven $\Dm=\pm2$ and $\Dm=\pm1$ spin transitions, respectively. They agree well with the magnetic field strengths calculated from Eq.~(\ref{eq_Hamiltonian}) for the resonance frequency $\fSAW = 916$~MHz [cf.~double vertical arrows in Fig.~\ref{fig2}(a)]. Both resonances are well fitted by a Lorentzian function [solid curves in Fig.~\ref{fig2}(b)] with a full width at half maximum (FWHM) of 2.2~mT and 6.0~mT. Note that the resonances are actually doublets, which are split by approximately $1$~mT and $2$~mT for the $\Dm=\pm2$ and $\Dm=\pm1$ spin transitions, respectively. These splittings are not resolved due to a relatively large broadening caused by a reduction of the spin coherence in proton-irradiated samples with high irradiation fluences \cite{Kasper:2020}.

A remarkable property of the SAR interaction illustrated in Fig.~\ref{fig2}(b) is the ability to selectively couple all spin states within the $\VSi$ qudit \footnote{In large magnetic fields considered here, the spin-acoustic interaction does not couple the $m_S = -3/2$ and $m_S = +3/2$ states. In case of ${\bf B} \sim D/\mu_B$, any pair within the four spin states can be coupled acoustically.}. In particular, the $\Dm=\pm2$ transitions are normally forbidden for RF-driven spin resonance. Therefore, their excitation in Fig.~\ref{fig2}(b) represents a clear evidence of the acoustic nature of the observed resonances. To further corroborate this acoustic nature, we display in Fig.~\ref{fig2}(c)  the intensity of the optically detected $\Dm=\pm2$ SAR as a function of the magnetic field strength (vertical axis) and the RF frequency applied to the SAW resonator (horizontal axis). The SAR signal vanishes as soon as either $B$ is changed or $\fSAW$ is detuned (cf. $\sleft$ in the same panel), thus confirming that the spin transitions are caused by the dynamic fields of the SAW. Additional studies summarized in the SM \cite{Note2} show that the spatial dependence of the SAR intensities follow the distribution of the acoustic field within the SAW resonator. We also prove that our experiments are performed in the linear regime for all observed SARs \cite{Note2}. 

\begin{figure}
\includegraphics[width=\linewidth]{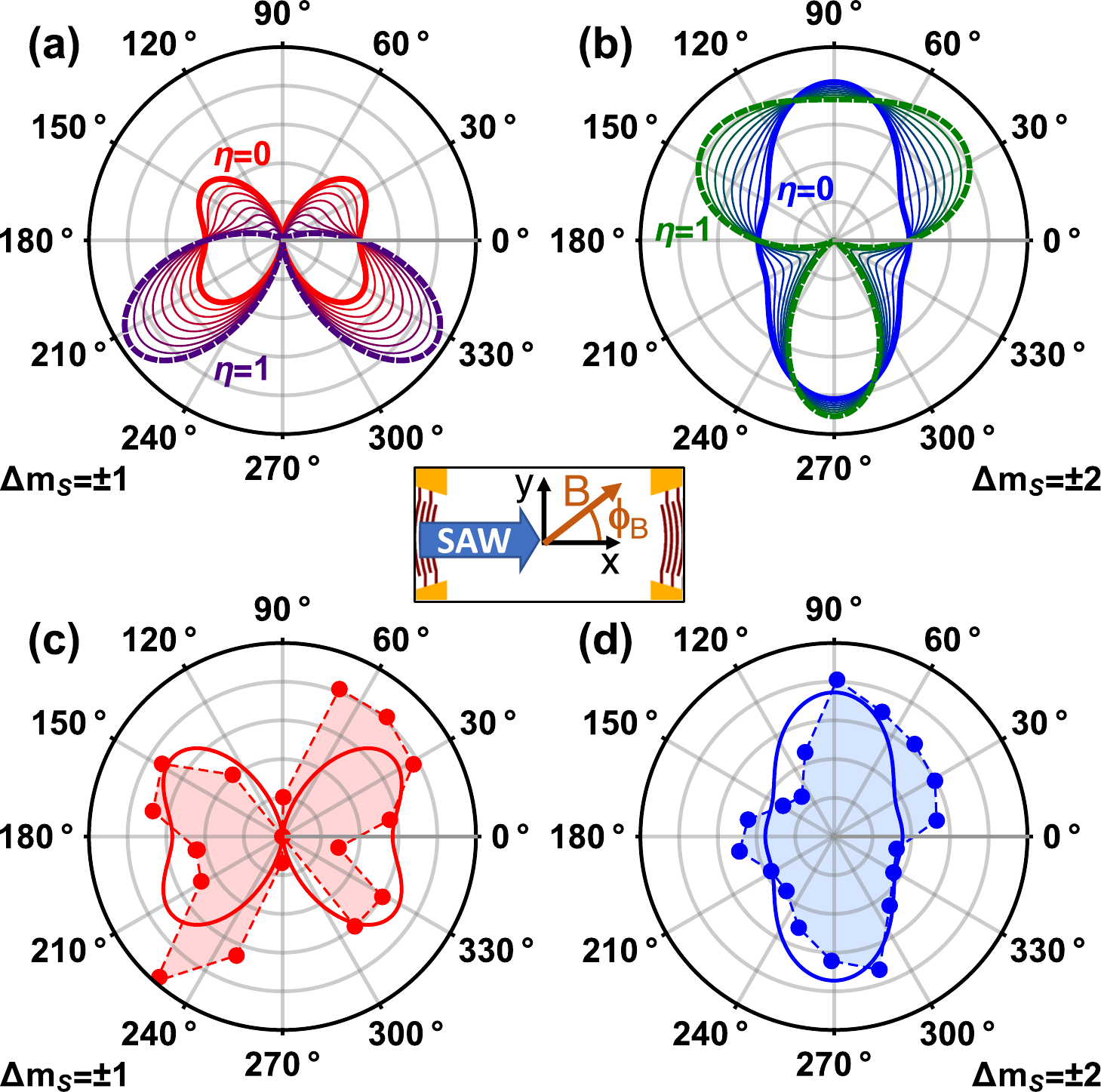}
\caption{(a,b) 
Dependence of the $\Dm=\pm 1$ and $\Dm=\pm 2$ transition intensities on the angle $\phi_B$ between the in-plane magnetic field and the SAW propagation direction (see inset) calculated for the cases of standing wave ($\eta=0$, thick thick lines), traveling wave ($\eta=1$, dashed thick lines), and intermediate values of $\eta$. The strain  components used are $\langle u_{xx}^2 \rangle = 9.0 \times 10^{-9}$, $\langle u_{zz}^2 \rangle = 7.0\times 10^{-9}$, $\langle u_{xz}''^2 \rangle = 3.5\times 10^{-9}$ , $\langle u_{xx} u_{zz} \rangle = 7.7\times 10^{-9}$, $\langle u_{xx} u_{xz}'' \rangle = 
 5.3\times 10^{-9}$, $\langle u_{xz}'' u_{zz} \rangle = 5.0\times 10^{-9}$, as obtained from the distributions of Fig.~\ref{fig2}(d). (c,d)  Angular dependencies of the $\Dm=\pm 1$ and $\Dm=\pm 2$ transition amplitudes as a function of the in-plane magnetic field orientation with respect to the SAW propagation direction. Circles present the experimental data, the thin dashed lines are guides to the eye. The thick lines show the calculation after Eq.~\eqref{eq:Ma} with  $\eta=0$. }
\label{fig3}
\end{figure}

We are now in the position to discuss the anisotropic nature of the spin-acoustic resonances, which is a further important finding of this work. We assume the reference frame illustrated in Fig.~\ref{fig1}(b) with the SAW beam propagating along the $x$ axis. Figure~\ref{fig2}(b) compares the optically detected SAR signal for two angles $\Bangle$ between the SAW propagation direction and the in-plane magnetic field. While the magnetic field strengths at which the SARs take place are independent of the in-plane orientation of $\vBext$, the SAR intensities do depend clearly on $\Bangle$. The full anisotropic behavior with respect to the field  orientation on the sample plane is summarized by the circles in Figs.~\ref{fig3}(c) and \ref{fig3}(d), which display  the intensity of the $\Dm=\pm1$ and $\Dm=\pm2$ SARs, respectively, as a function of $\Bangle$.

To understand the unusual angular dependence of the SARs, we develop a microscopic model for the spin-acoustic interaction. In the spherical approximation, the effect of a lattice deformation on a spin center is described by the interaction term   
\begin{align}
\mathcal{H}_{\rm def} = \Xi \sum_{\alpha\beta} u_{\alpha\beta} S_\alpha S_\beta \,,
\end{align}
\noindent where $\Xi$ is the interaction constant~\cite{Poshakinskiy:2019bi}. Being quadratic in the spin operators, such an interaction can induce spin transitions with  $\Dm=\pm 1$  as well as with $\Dm=\pm 2$. For ${\bf B} \parallel x$, the spin transitions with $\Dm=\pm 1$ are induced by the strain tensor components $u_{xy}$ and $u_{xz}$, while those with $\Dm=\pm 2$ are induced by the other linear-independent components $u_{yy}$, $u_{zz}$ and $u_{yz}$. The strain components responsible for the spin transitions for other ${\bf B}$ directions can be obtained by the corresponding rotation of the strain tensor. 

A  plane Rayleigh SAW propagating along $x$ is described by the strain tensor

\begin{equation}
\label{eq:wave}
u_{\alpha \beta} (t,x,z) = u_{\alpha \beta}(z) \e^{ikx-i\omega t} + u_{\alpha \beta}^*(z) \e^{-ikx+i\omega t}
\end{equation}

\noindent with non-vanishing components $u_{xx}$, $u_{zz}$, and $u_{xz} = \rmi u_{xz}''$~\cite{Rayleigh_PLMSs1-17_4_85}. We assume a reference frame for which $u_{xx}$, $u_{zz}$, and $u_{xz}''$ are purely real. The factor $\rmi =\sqrt{-1}$  indicates that the phase of the $u_{xz}$ component is shifted by $\pi/2$, thus resulting in an elliptically polarized strain field in the $xz$ plane. Figure~\ref{fig2}(d) compares the calculated depth profiles of the $\epsxx$, $\epszz$ and $\epsxz$ strain components \cite{PVS156} with the simulated depth distribution of the $\VSi$ defects \cite{SRIM}.

In our case, spin centers are inserted in an acoustic resonator and thus subject to a combination of two counterpropagating SAWs travelling along $x$ and $-x$ with intensities $I_+$ and $I_-$, respectively.  We use the parameter $\eta = (I_+ - I_-)/(I_++I_-)$ to distinguish different situations: a standing wave ($|\eta| = 0$), a traveling wave ($|\eta| = 1$) or intermediate cases ($0 < |\eta| < 1$). The rates $W_{\pm 1}$ and $W_{\pm 2}$ of the spin transitions with  $\Delta m_S=\pm 1$ and $\Delta m_S=\pm 2$, respectively, are then given by (see SM \cite{Note2}) 

\begin{align}\label{eq:Ma}
W_{\pm 1} &\propto 3 \cos^2\phi_B \langle u_{xx}^2 \sin^2\phi_B +u_{xz}''^2 +  2 \eta u_{xx}u_{xz}'' \sin\phi_B \rangle  \nonumber \\
W_{\pm 2} &\propto \frac{3}4 \langle (u_{xx}\sin^2\phi_B - u_{zz} )^2 +  4 u_{xz}''^2\sin^2\phi_B  + \nonumber\\
 &\qquad 4\eta ( u_{xx}\sin^2\phi_B - u_{zz}  ) u_{xz}''\sin\phi_B \rangle \,.
\end{align}
\noindent 

\noindent The transition rates in Eq.~\eqref{eq:Ma} were averaged along $x$ to account for the finite detection spot size, which is  larger than the SAW wavelength. The angular brackets $\langle \rangle$ indicate averaging along $z$ to take into account the depth distribution of the $\VSi$ centers as well as the strain field, as presented in Fig.~\ref{fig2}(d).

Finally, we analyze the symmetry of the SARs. Figs.~\ref{fig3}(a)-(b) present the angular dependencies of the $\Dm=\pm 1$ and $\Dm=\pm 2$ transition intensities, respectively, calculated after Eq.~\eqref{eq:Ma} for various $\eta$. The SARs are always symmetric with respect to the inversion of the $B_x$ component, since our system has a $(xz)$ mirror plane. For $\eta=0$, the SARs are also symmetric with respect to the inversion of the $B_y$ component due to additional presence  of the time-reversal symmetry. As $|\eta|$ increases, the latter symmetry  breaks as the strain field of the travelling SAW acquires an elliptical polarization. Particularly, Fig.~\ref{fig3}(a) shows that the $\Dm=\pm 1$  SAR almost vanishes for $B_y>0$ ($0^\circ < \Bangle < 180^\circ$) while it remains strong for $B_y<0$ ($180^\circ < \Bangle < 360^\circ$). Such an asymmetric angular dependence is a clear evidence of  the broken time-reversal symmetry in the presence of a travelling SAW. Upon inversion of the SAW propagation direction ($\eta<0$, not shown), the angular dependencies of such chiral SAR are flipped with respect to the horizontal axis. 

Having developed a microscopic model for the anisotropic SAR, we now apply it to analyze the  experimental data given by the circles in Figs.~\ref{fig3}(c) and \ref{fig3}(d). The angular dependence of the $\Dm=\pm1$ SAR has a butterfly-like shape with vanishing signal for $\Bangle=\pm90^\circ$ and maxima when the magnetic field rotates towards $\Bangle \approx \pm45^\circ$ or $ \pm135^\circ$. In contrast, the angular dependence of the $\Dm=\pm2$ SAR has a cocoon-like shape with maxima for $\Bangle=\pm90^\circ$. This SAR  does not vanish for any direction of the in-plane magnetic field. These measured angular dependences are best reproduced by  Eq.~\eqref{eq:Ma} by assuming  $\eta=0$, which yields the solid lines in Fig.~\ref{fig3}(c) and \ref{fig3}(d). This result is consistent with the expected standing-wave nature of the acoustic fields within a resonator. We emphasize that our model has no fitting parameters except for the overall intensity to match the readout optical signal.

In conclusion, we observe half-integer SAR in SiC at room temperature. Using a SAW resonator patterned on the SiC surface, we are able to address both the $\Dm=\pm 1$ and the $\Dm=\pm 2$ spin transitions of the $\VSi$ spin-3/2 center with all-optical readout and without requiring extra microwave electromagnetic fields. The SARs reveal a complex behavior, which depends on the magnetic field orientation with respect to the SAW propagation direction. Our theoretical model describes these angular dependencies without any fitting parameter and predicts chiral spin-acoustic interaction for traveling SAWs. Such a room-temperature hybrid spin-mechanical platform can be used to implement quantum sensors \cite{Kraus:2013vf} with on-chip SAW control instead of microwave electromagnetic fields \cite{Poshakinskiy:2019bi} as well as to realize acoustically driven topological states \cite{PRL.Chiral}. 

\begin{acknowledgements}
The authors would like to thank S.~Meister and S.~Rauwerdink for technical support in the preparation of the samples, S.~A.~Tarasenko and M.~Helm for discussions and critical questions, and S. Fölsch for a critical reading of the manuscript. A.V.P. acknowledges the support  from the Russian Science Foundation (project 20-42-04405) and the Foundation ``BASIS''. G.V.A. acknowledge the support from  the German Research Foundation (DFG) under Grant  AS 310/5-1. Support from the Ion Beam Center (IBC) at Helmholtz-Zentrum Dresden-Rossendorf (HZDR) is gratefully acknowledged for the proton irradiation.
\end{acknowledgements}


%

\end{document}


\title{Anisotropic Spin-Acoustic Resonance in Silicon Carbide at Room Temperature}
\author{A.~Hern\'{a}ndez-M\'{i}nguez}
\email{alberto.h.minguez@pdi-berlin.de}
\affiliation{Paul-Drude-Institut f\"{u}r Festk\"{o}rperelektronik, Leibniz-Institut im Forschungsverbund Berlin e.V., Hausvogteiplatz 5-7, 10117 Berlin, Germany}
\author{A.~V.~Poshakinskiy}
\affiliation{Ioffe Physical-Technical Institute, Russian Academy of Sciences, 194021 St.~Petersburg, Russia}
\author{M.~Hollenbach}
\affiliation{Helmholtz-Zentrum Dresden-Rossendorf, Institute of Ion Beam Physics and Materials Research, Bautzner Landstrasse 400, 01328 Dresden, Germany}
\affiliation{Technische Universit\"{a}t Dresden, 01062 Dresden, Germany}
\author{P.~V.~Santos}
\affiliation{Paul-Drude-Institut f\"{u}r Festk\"{o}rperelektronik, Leibniz-Institut im Forschungsverbund Berlin e.V., Hausvogteiplatz 5-7, 10117 Berlin, Germany}
\author{G.~V.~Astakhov}
\affiliation{Helmholtz-Zentrum Dresden-Rossendorf, Institute of Ion Beam Physics and Materials Research, Bautzner Landstrasse 400, 01328 Dresden, Germany}


\begin{abstract}
This Supplemental Material contains information about:
\begin{itemize}
\item Photoluminescence characterization
\item ODMR spectrum at zero magnetic field
\item Spatial dependence of the spin-acoustic resonance
\item Dependence of the spin-acoustic resonance on SAW power
\item Matrix elements of the SAW-induced spin transitions
\end{itemize}
\end{abstract}

\maketitle

\section{Photoluminescence characterization}\label{sec_PL}

\begin{figure}[h]
\includegraphics[width=0.8\linewidth]{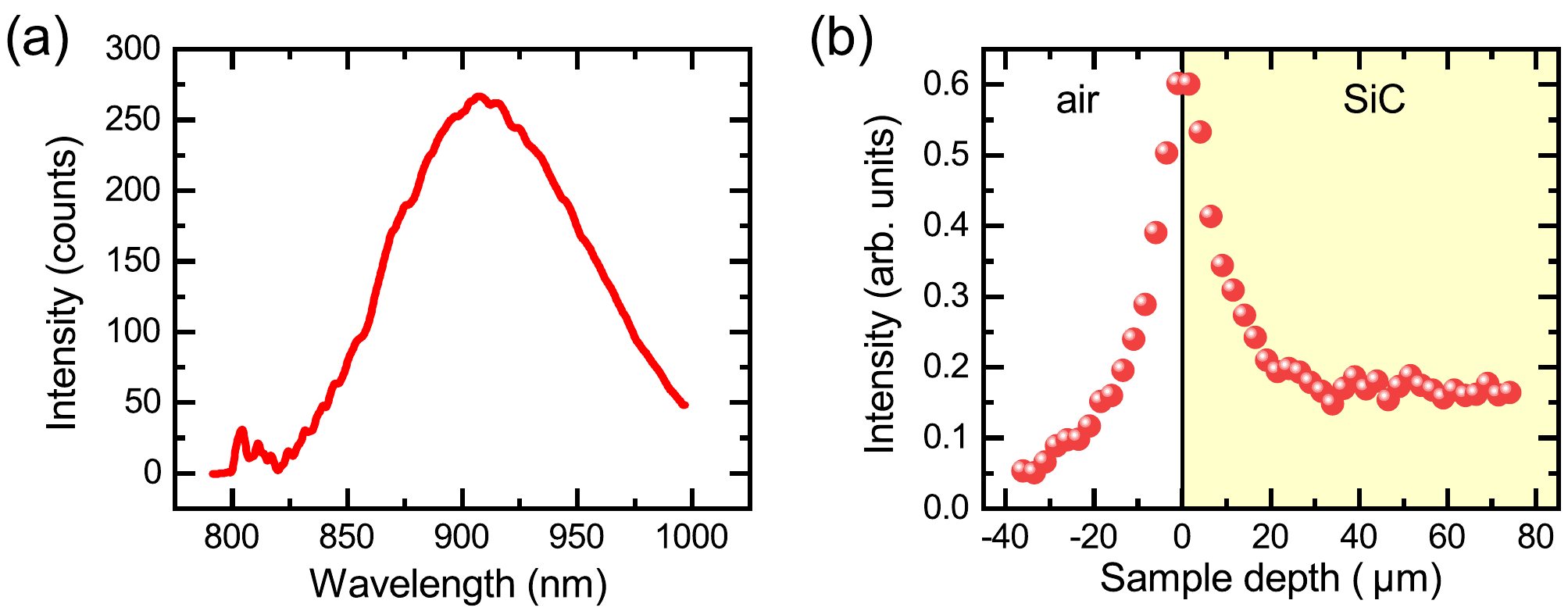}
\caption{(a) Room-temperature PL spectrum of the $\VSi$ ensemble in 4H-SiC. (b) Depth profile of the PL intensity across the surface of the 4H-SiC.}
\label{figS1}
\end{figure}

Figure~\ref{figS1}(a) displays the typical room-temperature photoluminescence (PL) spectrum of the $\VSi$ ensemble after proton irradiation of the 4H-SiC. Figure~\ref{figS1}(b) shows the depth profile of the PL intensity. The $\VSi$ defects implanted at a depth of 250~nm cause the strongest PL signal. The measurements are taken in the same confocal setup as for the spin-acoustic resonance (SAR) experiments.

\newpage

\section{ODMR spectrum at zero magnetic field}\label{sec_ODMR}

\begin{figure}[h]
\includegraphics[width=0.4\linewidth]{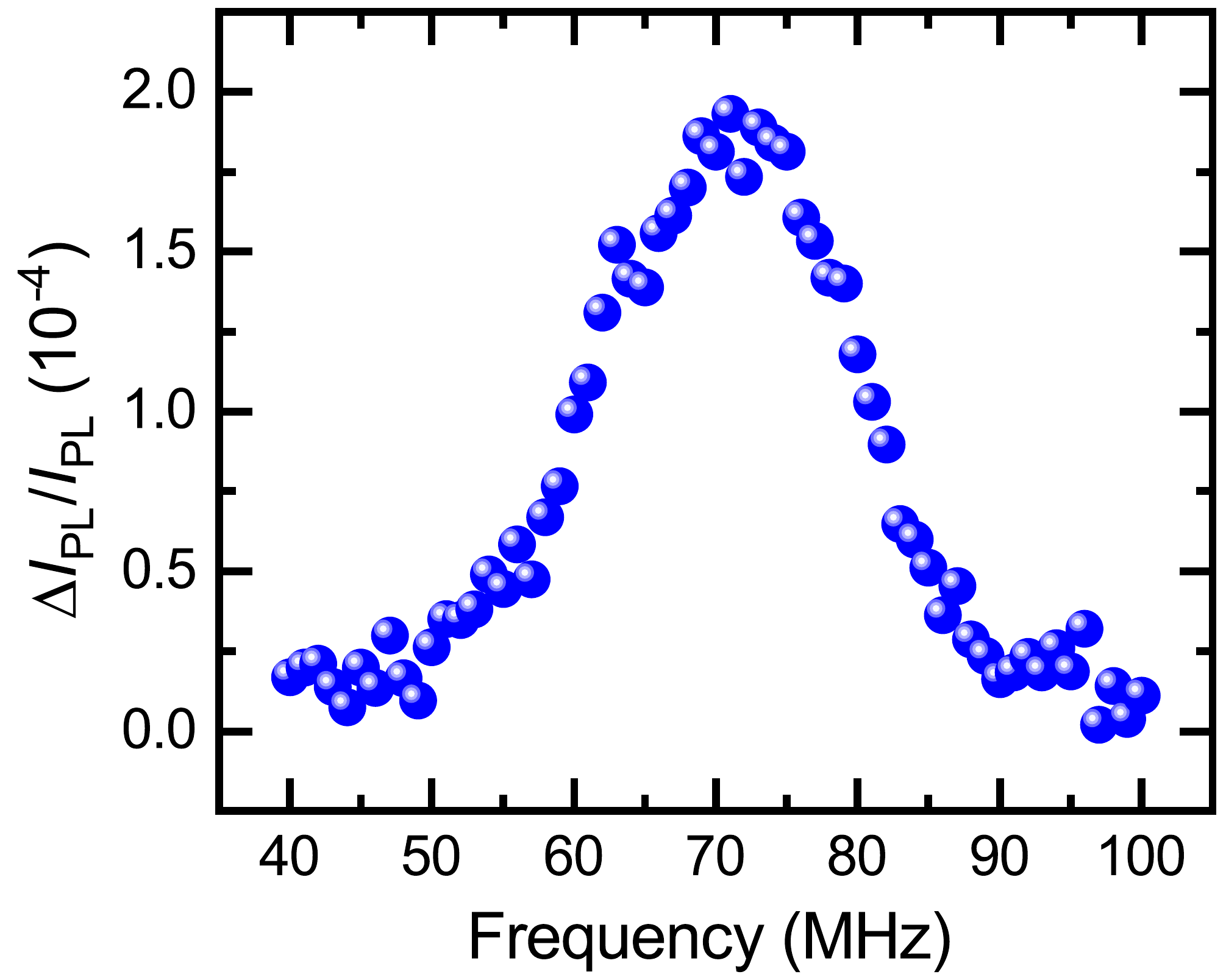}
\caption{Room-temperature ODMR spectrum at zero magnetic field of the $\VSi$ ensemble implanted at the surface of the 4H-SiC.}
\label{figS2}
\end{figure}

Figure~\ref{figS2} displays the optically detected magnetic resonance (ODMR) spectrum of the $\VSi$ ensemble at zero magnetic field. The measurements are taken at room temperature in the same confocal setup as for the SAR experiments. The microwave frequency is applied by a coplanar waveguide placed close to the sample. The 70~MHz peak corresponds to the splitting frequency between the $\mS=\pm1/2$ and $\mS=\pm3/2$ spin sublevels of the ground state.

\section{Spatial dependence of the spin-acoustic resonance}\label{sec_yscan}

\begin{figure}[h]
\includegraphics[width=0.8\linewidth]{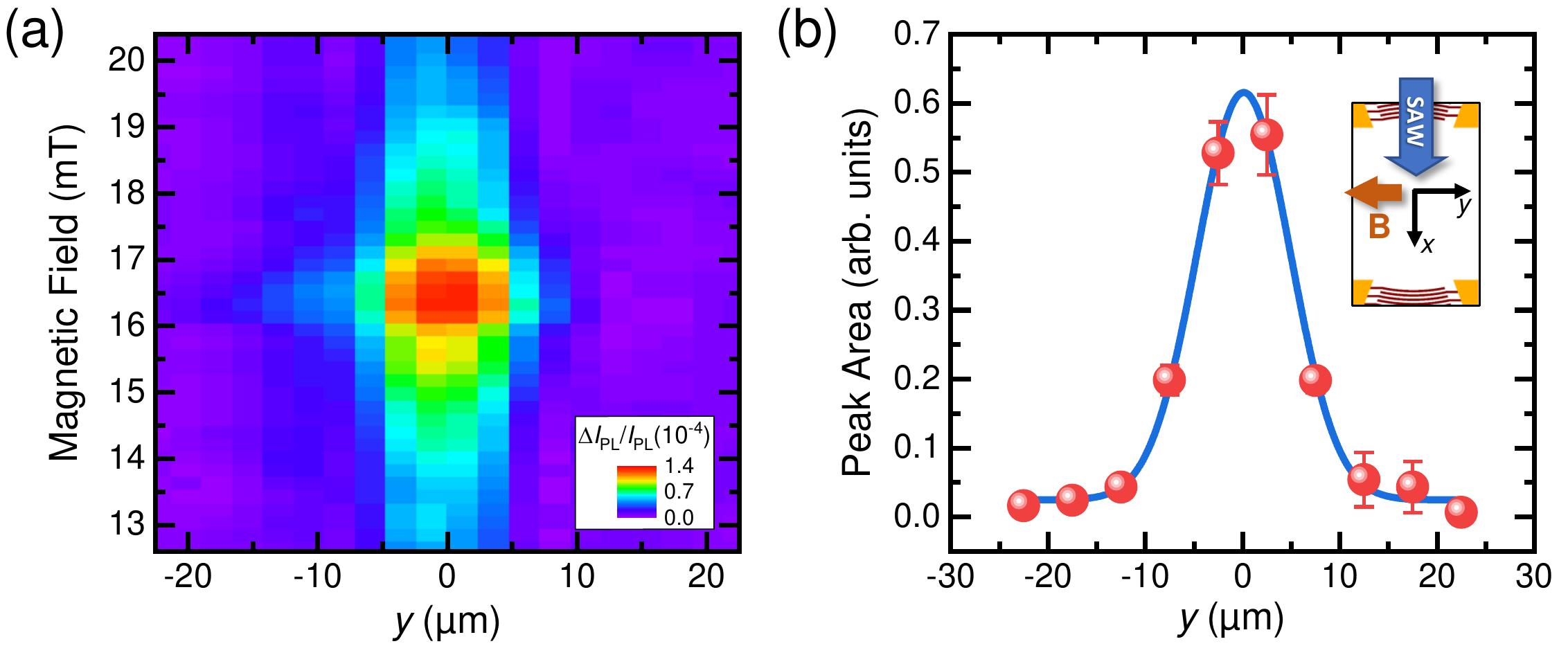}
\caption{(a) Two-dimensional color map of the $\Dm=\pm2$ SAR as a function of magnetic field strength and position of the exciting laser across the center of the SAW resonator. (b) Area of the $\Dm=\pm2$ peak as a function of the laser position across the center of the SAW resonator. The blue curve is a Gaussian fit to the data.}
\label{figS3}
\end{figure}

The colormap of Fig.~\ref{figS3}(a) displays the intensity of the optically detected $\Dm=\pm2$ SAR as a function of the magnetic field strength (vertical axis) and position of the exciting laser across the center of the SAW resonator (horizontal axis). The SAR signal vanishes as soon as the laser moves away from the center of the resonator, defined as $y=0$. Figure~\ref{figS3}(b) shows the amplitude of the resonance, defined as the area of the signal in Fig.~\ref{figS3}(a) when integrated along the magnetic field, for several positions of the laser. The spatial dependence of the SAR signal is well fitted to a Gaussian peak with a FWHM of $11~\um$. This is about three times narrower than the finger length at the inner part of the IDTs  ($35~\um$), thus confirming the focusing efficiency of our acoustic resonator and further supporting the acoustic nature of the SAR signal.

\newpage

\section{Dependence of the spin-acoustic resonance on SAW power}\label{sec_SAWpower}

\begin{figure}[h]
\includegraphics[width=0.8\linewidth]{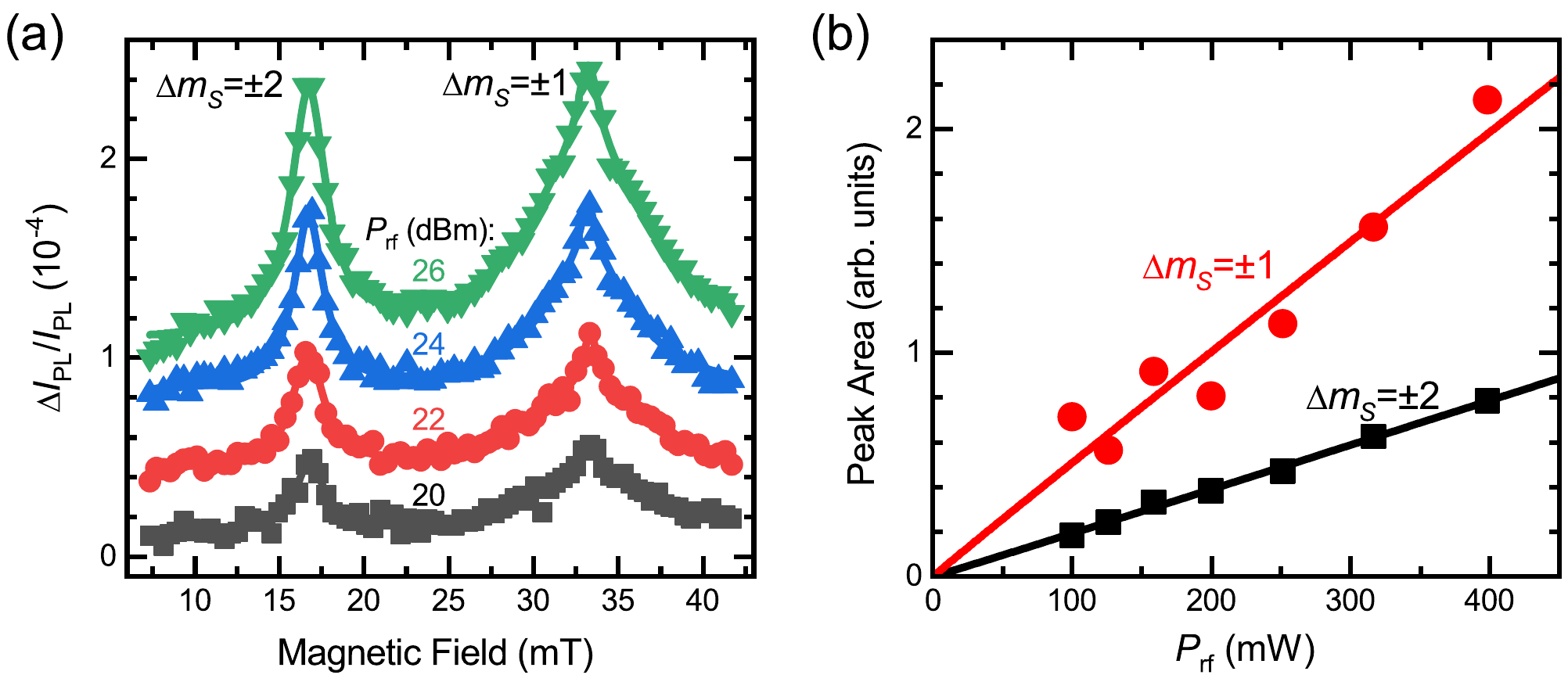}
\caption{(a) Optically detected SAR signal as a function of the magnetic field strength, measured for several nominal RF powers $\Prf$ applied to the acoustic resonator. The data are taken for an angle $\Bangle=30^\circ$ between the in-plane magnetic field and SAW propagation direction. The solid curves are Lorentzian fits to the data. The data are vertically shifted for clarity. (b) Area of the SAR peaks as a function of $\Prf$. The solid curves are linear fits to the data.}
\label{figS4}
\end{figure}

Figure~\ref{figS4}(a) displays the dependence of the optically detected SAR signal on the nominal RF power applied to the acoustic resonator, $\Prf$. Figure~\ref{figS4}(b) displays the area of both resonances [obtained from Lorentzian fits to the data in Fig.~\ref{figS4}(a)] as a function of $\Prf$, together with their linear fits. 

\section{Matrix elements of the SAW-induced spin transitions}\label{sec_theory}

First, we choose the new reference frame $(x'y'z)$ where the new in-plane axes are $x' \parallel \bf B$ and $y' \perp \bf B$. In such frame, the matrix elements $M^{(+)}_{\Delta m_S}$ of spin transitions with the change of the spin projection on $x'$ axis $\Delta m_S$ induced by absorption of the SAW read
\begin{align}
M^{(+)}_{\pm 1} &= \sqrt3 \Xi (u_{x'y'} \mp \rmi u_{x'z}) 
\,,\\ \nonumber
M^{(+)}_{\pm 2} &= \tfrac{\sqrt3}2 \Xi(u_{y'y'}-u_{zz} \mp 2\rmi u_{y'z}) 
\,.
\end{align}
To obtain the matrix elements for the transitions with emission of the phonon $M^{(-)}_{\Delta m_S}$, one must replace $u_{\alpha\beta}$ with $u^*_{\alpha\beta}$ in the above equation. Expressing the strain components in the new basis via those in the original basis $u_{xx}$, $u_{zz}$, and $u_{xz}$, we find
\begin{align}
M^{(+)}_{\pm 1} &= \sqrt3 \Xi \cos\phi_B ( u_{xx} \sin\phi_B \mp \rmi u_{xz} )  
\,,\\ \nonumber
M^{(+)}_{\pm 2} &= \tfrac{\sqrt3}2 \Xi ( u_{xx}\sin^2\phi_B  - u_{zz} \mp 2\rmi u_{xz}\sin\phi_B)
\,.
\end{align}
We note that transitions with $\Delta m_S >0$ correspond to an increase of energy and are accompanied by SAW absorption, while $\Delta m_S <0$ correspond to a decrease of energy and are accompanied by the emission of SAW quanta. Therefore, the relevant matrix elements are $M^{(\pm)}_{\pm 1}$ and $M^{(\pm)}_{\pm 2}$. We consider that the strain field is a sum of two waves of the form $u_{\alpha \beta} (t,x,z) = u_{\alpha \beta}(z) \e^{ikx-i\omega t} + u_{\alpha \beta}^*(z) \e^{-ikx+i\omega t}$ with wave vectors $k$ and $-k$ and different intensities $I_+$ and $I_-$. Note that the $u_{xx}$ and $u_{zz}$ strain components of the right- and left-traveling waves match while the $u_{xz}$ component is opposite. The spin transition rates are obtained by averaging $W_{\pm 1} = |M^{(\pm)}_{\pm 1}|^2$ and $W_{\pm 1} = |M^{(\pm)}_{\pm 2}|^2$ over the $x$ and $z$ directions.